# A Proposed Architecture for Big Data Driven Supply Chain Analytics

Sanjib Biswas[1] and Jaydip Sen[2]

[1]Calcutta Business School, Bishnupur- 743503, South 24 Parganas, West Bengal.
E-mail: sanjibb@acm.org

[2]Calcutta Business School Bishnupur- 743503, South 24 Parganas, West Bengal
Email: jaydip.sen@acm.org

**Abstract**

Advancement in *information and communication technology* (ICT) has given rise to explosion of data in every field of operations. Working with the enormous volume of data (or Big Data, as it is popularly known as) for extraction of useful information to support decision making is one of the sources of competitive advantage for organizations today. Enterprises are leveraging the power of analytics in formulating business strategy in every facet of their operations to mitigate business risk. Volatile global market scenario has compelled the organizations to redefine their *supply chain management* (SCM). In this paper, we have delineated the relevance of Big Data and its importance in managing end to end supply chains for achieving business excellence. A Big Data-centric architecture for SCM has been proposed that exploits the current state of the art technology of data management, analytics and visualization. The security and privacy requirements of a Big Data system have also been highlighted and several mechanisms have been discussed to implement these features in a real world Big Data system deployment in the context of SCM. Some future scope of work has also been pointed out.

## 1. Introduction

In recent years, SCM has become one of the key enablers for achieving competitive advantage. An effective SCM can be proved critical for the success or failure of an organization and thus it becomes an important value driver for organizations. Increased customer demand and variety, intensified competition, increasing complexity and dynamicity of global operations, pressure on innovation of products and services, advances in technology particularly *information and communication technology* (ICT) have added complexity in designing and managing supply chains. Over the years, the concepts and practices of SCM have undergone several changes that have been reflected in its 'constantly evolving' nature. From its initial cost efficiency focus to modern responsive and agile nature, SCM has witnessed a transformational change at the operational frontier. To sustain under volatile business environment, it has become imperative to operate with information driven strategies wherein collaboration among the members is one of the key success factors. Effective coordination and collaboration enables the members of a supply chain to achieve its global objectives.

However, sharing of information plays an indispensable role in SCM integration. It improves customer services and financial performances by providing accurate and relevant on-time information and also enhances supply chain visibility. It sets and monitors key performance indicators to highlight variances and inefficiencies and mitigates the *bullwhip effect* which is essentially caused due to the distortion of demand information while moving from downstream to upstream (Cheng et al., 2010; Lee & Whang, 2000; Lee et al., 1997; Miah, 2015; Vickery et al., 2003). How timely and accurately an organization can formulate an effective and futuristic strategy has become a critical issue in the context of modern SCM. Exploiting the rich capabilities of analytics, organizations can reap the benefits of Big Data-driven insights to work with optimal lead time and improve prediction of future to cope up with uncertainties. Researchers have found that in order to achieve seamless coordination or harmony among the members of a supply chain for taking right decision at right time, to deploy resources optimally and channelize all activities in right direction, to provide right product to the customers at right time, information acts as an *invisible thread* among the members.

With the rapid evolution and adoption of ICT by the industries, colossal amount of data is being generated all pervasively from every activity across a supply chain. According to the International Data Corporation (IDC), the estimated growth of digital data will be as high as 40 trillion gigabytes by 2020 as compared to approximately 2.8 trillion gigabytes in 2012. This opens up enormous opportunity for business organizations to effectively utilize such gigantic amount of data for making prudent business decisions. As the business world is progressing towards *Industrie 4.0*, every object linked with a supply chain is now acting as a continuous generator of data in structured or unstructured form. Germany Trade & Invest report (2014) described *Industrie 4.0*, the upcoming fourth industrial revolutions, as an intelligent ICT based decentralized real time production system. In this system, the real and virtual world are connected through a cyber-physical interface wherein the products and machineries independently exchange and respond to information for managing end to end processes. In essence, it offers a technological platform to agglutinate production technologies and smart processes to establish a smart factory (Germany Trade & Invest, 2014).It is providing organizations an unprecedented opportunity to leverage *informed* supply chain strategy for leveraging competitive advantage. Needless to mention, managing such gigantic volume of data - *Big Data*, as it is popularly known - is a stupendous task.

Understanding and tracking of data generation and then processing of data for deriving useful information to operate a smart supply chain stands as the key to success. Analytics thus plays a vital role in formulating smart strategies for enhancing the performance of a supply chain (Khan, 2013).

This paper is an extended work of our previous contribution (Biswas & Sen, 2016). In this paper, we have highlighted the importance of the role of information in integrating the components of a supply chain for formulating competitive strategies and predicting future changes. We have also emphasized the contextual significance of analytics in understanding and analyzing the data for extracting useful information which form the rationale for effective decision making. Here, we have summarized the evolution of supply chain analytics (SCA) and its tacit role in effective SCM and governance. In line with our previous work (Biswas & Sen, 2016) we propose an architecture for a Big Data-centric supply chain that is in conformance with the architectural standards. Several security and privacy requirements are highlighted and mechanisms to achieve them are also discussed.

The reminder of this paper is organized as follows. Section 2 discusses the evolution of SCA and its importance in the context of supply chain performance. In Section 3, the concept of Big Data and the relevance of Big Data analytics in context of SCM are pointed out. In Section 4, we have described a standard Big Data architecture in the literature and then have proposed a supply chain specific Big Data system. Section 5 presents various security and privacy issues that need to addressed in a Big Data system and also discusses numerous mechanisms and protocols for designing a secure Big Data architecture. Section 6 concludes the paper while highlighting some future scope of work.

## 2. Supply Chain Analytics (SCA) and its Impact on Supply Chain Performance

In his classical study, Christopher (1992) defined supply chain as "The network of organizations that are involved, through upstream and downstream linkages, in the different processes and activities that produce value in the form of products and services delivered to the ultimate consumer". The concept of SCM has continuously evolved from the initial concept of physical distribution in early 60's to enterprise level function over the last couple of decades. Lambert et al. (1998) represented SCM as a set of integrated key business processes initiated from end users and extended to ultimate suppliers which to provide products, services and information for adding value to the stakeholders. Traditionally, major focus was on *efficiency*, i.e. meeting demand with lowest possible cost. In the journey of evolution, SCM has witnessed a paradigm shift in its focus from cost reduction and serviceability (Christopher, 1998) to cost effectiveness, reliability and predictability of the future through various developments like *materials requirement planning* (MRP), *manufacturing resource planning* (MRP II), *enterprise resource planning* (ERP), *customer relationship management* (CRM), total quality leadership, *vendor managed inventory* (VMI), warehouse management and network planning, *collaborative planning, forecasting and replenishment* (CPFR) etc (Stock, 2013). Changing scenario of the global business environment with shorter product life cycles, availability of a gamut of product variety, shorter lead time from design to delivery while operating economically, have posed many challenges to the researchers and practitioners. SCM strategy must be an integral part of the overall corporate strategy and should be supported by appropriate organizational culture while establishing a strategic fit (Chopra & Meindl, 2012). In order to achieve such strategic fit, it is imperative to maintain *horizontal synergy* among the members of the supply chain (Yan et al., 2014). Experts believe that, it is the 'harmony' instead of 'competition' which determines the basis for futuristic SCM strategy.

Lambert & Cooper (2000) represented supply chain as a network of business and synergy in relationships across the chain in pursuit of business excellence. The authors identified the key issues in the context of SCM like consideration of key members, linked processes and the level of integration. The members have been classified into two categories: *primary* (responsible for carrying out value added activities for the customers) and *supporting* (facilitating primary functions by providing resources and knowledge). According to the authors, the key processes are designed on the horizontal or the vertical dimension to accomplish the functions of supply chain like customer service and relationship management, demand management, supply management, manufacturing and return management. The authors also argued that, conflict stems from inconsistencies in inter-organizational integration which results into supply chain inefficiencies (Lambert & Cooper, 2000).

Supply chains must be capable to promise rather than available to promise. Researchers have observed that order winning SCM strategy must support both efficiency in supplying functional products as well as responsiveness for supplying innovative products while maintaining agility (Fisher, 1997; Cheung et al., 2012). Agility infuses dynamicity in the system and escalates the capability of the supply chain to respond and attune to unforeseen changes while operating in a volatile and highly competitive business environment. In their independent studies, Thompson (1967) and Drucker (1968) ascertained that, one of the most important tasks for an enterprise is to search or anticipate for changes, accept and respond to the changes for exploring opportunities to mitigate the effect of uncertainty. Azvine et al. (2007) argued that survival and prosperity of an organization depends on how well the organization understands and responds to the changing environment. An agile organization is able to sense the changes, predict the dimension and nature of change and perceive the change in its true sense. Razmi & Ghasemi (2015) observed that trust plays a vital role in mediating between organizational commitment and technological capabilities.

Sharifi & Zhang (2001) reported the importance of agility in the context of achieving sustainability in business while operating in a turbulent environment. The authors opined that adaptability to change and adoption of proactive strategies to derive benefits out of the changes act as the basis for agility. In line with the prior studies in the literature, the authors represented agility as the integration of technology, process, people and organization operating with flexibility and adaptability. The authors have also put forward a model for agility and a framework for its validation. The model stands on three constituting elements: (i) *agility driver* (need for change exerted by the environmental elements like response time, variation in needs and preferences, reduction in time to market, higher quality expectation, process capability enhancement, innovation etc.), (ii) *agility capability* (ability of the organization such as responsiveness, skill and flexibility, to quickly respond to the changes for extracting benefits) and (iii) *agility providers* (organizational elements like technology, people, process and innovation which establish the capabilities) (Sharifi & Zhang, 2001).

In essence, integration becomes a vital factor for effective functioning of supply chains to meet the performance objectives like cost, responsiveness, serviceability, agility etc. Research has revealed that the effectiveness of resource − performance relationships linked with global objectives across the supply chain depend significantly on the mutual collaboration and coordination among the partners, which in turn facilitates effective decision making and formulation of effective supply chain strategies (Dong et al., 2009; Sahin & Robinson, 2002; Cheng et al., 2010; Van Donk & Van der Vaart, 2005; Power, 2005). In other words, SCM is the strategic coordination among the members of the supply chain to integrate supply and demand management (Stevenson, 2009). This highlights the importance of integrating both internal and external systems i.e. the *arc of integration* extends from the supplier end to the customer end through the periphery of the focal system (Salo & Karjaluoto, 2006; Frohlick & Westbrook, 2001).

Essentially according to the *supply chain operations reference* (SCOR) framework, the integration of demand and supply management across a supply chain takes place through four broad processes like plan, source, make and delivery, which involves flow of material, information and fund. Performance of such a system depends on planning, target setting, monitoring and control (Cai et al., 2009). In the context of SCOR model, Hoole (2005) put forward the competitive priorities of a supply chain. An effective SCM intends to provide consistent quality in response to the market needs with agility while eliminating waste out of the system to reduce cost and generate supply chain surplus. However, it requires access to information on time and an efficient data management for governing the activities and performances and hence having business insight to explore opportunities and overcome difficulties (Hoole, 2005).

It is evident from the above discussion that, *information technology* (IT) has become instrumental in the formulation of competitive supply chain strategy today. Ryssel et al.(2004) interpreted IT as a technology which enables to communicate, interpret, exchange and use information in the forms of data, voice, images, videos. Reddi & Snow (1993) contend that IT reduces coordination cost and transactional risk that leads to better governance. They further argued that inter-firm coordination and collaboration with the help of IT does not largely depend on the hierarchical structure and nature of relationship. Thus, unlike traditional method of coordination, investment in IT always pays off (Reddi & Snow, 1993). Salo & Karjaluoto (2006) accentuated the importance of IT in bringing transparency in supply chain operations. They advocated for an active involvement of the top management for strategic use of IT in SCM. The authors mentioned that in order to successfully implement IT for improving supply chain performance connectivity among relevant people and organizations is crucial. The authors further noted that there must be collaboration among both the virtual value chain and the physical value chain for providing required value to the end customers (Salo & Karjaluoto, 2006). However, they have also pointed out that in order to avail the benefits out of integration and operating with transparent supply chains, information must be accurate, precise and timely shared. Liker & Choi (2004) commented that, "…sharing a lot of information with everyone ensures that no one will have the right information when it's needed (p.112)". What's really required is the availability of timely, accurate, precise and relevant information for effective decision making. For example, to optimize inventory in a supply chain, precise demand information is necessary otherwise it results into unmatched inventory level. Consequently, the efficiency of the supply chain gets affected significantly resulting into high opportunity cost.

Information in the context of SCM broadly includes customer information, sales information, market and competitor information, product and service level requirement, promotion/brand information, demand forecasting, inventory, capacity utilization, process planning and control information, skill inventory, human information, sourcing/vendor information, networking information, logistics, warehouse planning, pricing and fund flow/working capital information. Therefore, the role of data can never be over emphasized in the context of SCM. We highlight the critical role of data in supply chain in Figure 1 (Biswas & Sen, 2016). It depicts a typical data driven supply chain structure. In this structure, demand is initiated at customer end which flows through subsequent stages to the supplier end. Supply of goods and services follows the path from the supplier end to the customer end. Return of goods for repairing, remanufacturing and recycling follows the reverse path to that of demand. The data being generated at subsequent stages are classified broadly into four categories based on the type of the process and nature of usage. Supplier data essentially are linked with the activities pertaining to sourcing process. Manufacturing data is generated out of the conversion activities performed by the manufacturer. After manufacturing, the product is being delivered to warehouses from where it is distributed to the end customers. Delivery data provides the delivery information. Sales and distribution data contains customer information related to sales and product demand. Table 1 exhibits different types of data generated at different stages of a typical supply chain (Biswas & Sen, 2016).

**Table1.** Types of data generated at different stages of a supply chain

| Node | Data generation |
|---|---|
| Supplier | Design data, Order status, Stock level, Schedule, Shipment & Routing, Return/Dispose, Finance data (e.g., a/c receivable, tax, pricing etc.) |
| Mfg. | Basic/Activity data, Design data, forecasting data, Prod. Plan/schedule, Capacity planning data, Process data (Lot size, cycle time, takt time, throughput time, process capability etc.), Yield data, Quality/Reliability data (FTR, % rejection, % failure etc.), Stock (RM/WIP/FG), Maintenance records, Customer feedback data, Vendor data, People data, Finance data (wage, conversion cost etc.), Return/dispose |
| Warehouse/ Distributor/ Retailer | Demand, Stock level, Schedule, Shipment & Routing, Order, Return/Dispose, Customer feedback, Finance data (pricing, payment etc.) |
| Customer | Point of sales (POS), Order status/ Demand, Product feedback, Customer opinions, Payment, Delivery, New product, Promotion/Recommendation, Return/ Dispose |

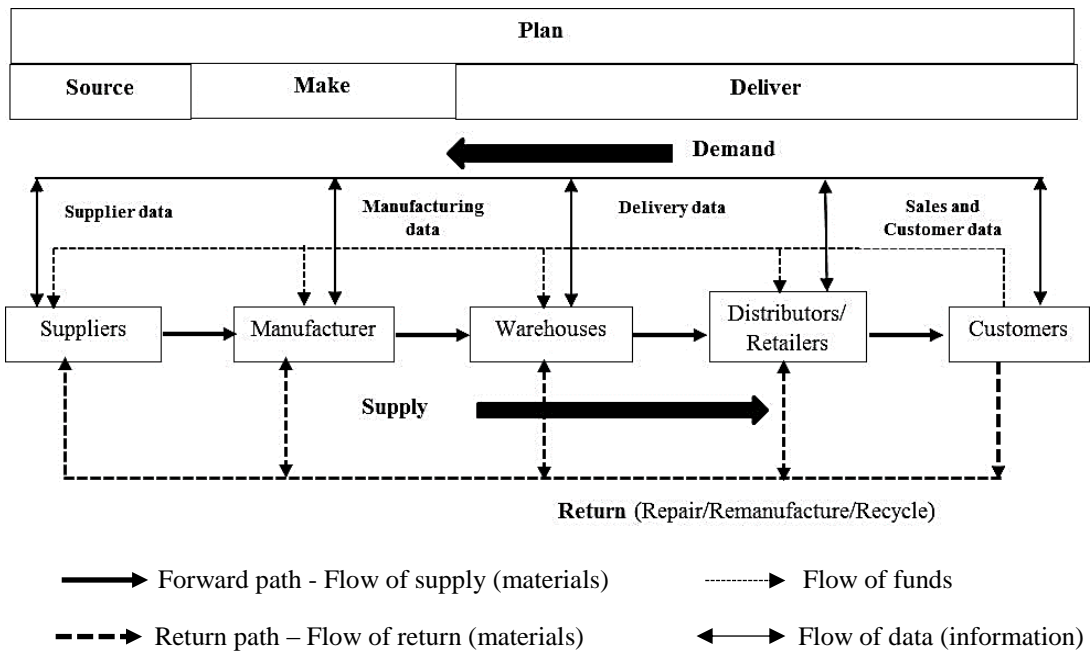

**Figure1**. Data driven general supply chain structure

However, growing ubiquitous nature of mobile and social communications and magnification of sensors have brought about surge in generation of data of diversified nature from multiple sources across an 'intelligent' supply chain. The study of Sahay & Ranjan (2008) highlighted the importance of real time *business intelligence* (BI) for improving the efficiency and effectiveness of a metrics-driven and fact based SCM. The authors interpreted BI as a technology platform based on advanced data mining techniques and algorithms which is used for data analysis to support decisions making. The authors also highlighted the importance of integrating a real time BI with the basic management functions for understanding and controlling the operations in a better way and for predicting the future requirements. For example, predictive analytics can be used to identify and manage inventory ageing and product returns for deriving best possible economic value (Sahay & Ranjan, 2008). Azvine et al. (2007) posited that organizations need to have BI in place for tracking the status of the business continuously with respect to pre-set performance objectives for operating in an uncertain and dynamic business environment. BI consists of a wide range of analytical solutions for capturing, analyzing and providing useful information to support decision making. In essence, BI includes a data warehouse and a time based process monitoring and analysis system (Golfareelli et al., 2004). Gangadharan & Swamy (2004) further extended the definition of BI at enterprise level. They referred BI as an integrated architecture based on a large database and analytical abilities which have both operational and decision support applications. They included knowledge management, ERP, data mining and DSS into the traditional domain of BI (Gangadharan & Swamy, 2004). Real time BI has gained its importance since it delivers appropriate information with minimum latency of the order of few seconds after the actual event occurs. Unlike traditional information system, which is based on the analysis of historical data, real time information flow identifies and works on the problems or opportunities at the same pace with actual events. It influences many vital operational decisions like assessment of customer churn decisions, product opinions, predictive maintenance, transportation tracking etc.

The primary objective of a supply chain strategist is to extract useful information by analyzing humongous amount of data being generated from all the objects across the supply chain. However, the challenge lies in aggregating such huge diverse set of data generated from multiple sources and thereby providing on-time information to assess present situation and predict the future. Integrating the concepts of BI with supply chain operations, organizations build an intelligent SCM system which is capable of analyzing what-if scenarios and take smart decision. Hence, SCA is of paramount importance for enhancing dynamic capabilities of supply chains.

Davenport (2006) espoused analytics as 'the science of analysis'. However, in the survey report published by Deloitte & MHI (2014) SCA is defined as a set of "tools and techniques that harness data from a wide range of internal and external sources to produce breakthrough insights that can help supply chains reduce costs and risk whilst improving operational agility and service quality" (Deloitte & MHI 2014, p. 2). In this context, Ittmann (2015) mentioned that the top two strategic priorities of SCM are *supply chain analytics* (SCA) and multi-channel fulfilment which highlights the importance of the analysis of data churned out from the activities pertaining to SCM and multi-channel coordination. When it is regarding the value of analytical and evidence based winning strategy formulation, the example of an 'underrated basketball team' originally given by Lewis is noteworthy (Lewis, 2003). Sahay & Ranjan (2008) put forth the significance of SCA in a holistic way. According to them, SCA portrays a broad view of the total supply chain as it not only exhibits detail movement of products, fund and information across all the activities being carried out at different stages of a supply chain, but also it takes into account key result indicators and analytics to understand the process for taking corrective, preventive and predictive decisions. SCA enables the organizations to set up primary drivers for basic functioning of a supply chain which includes product and process planning, sourcing, manufacturing and forward and reverse logistics. SCA provides a performance linked futuristic approach for improving the efficiency of a supply chain in terms of cost optimization, operational efficiency, improved relationship among the members and delivery optimization. However, the authors also opined that defining business and customer requirements vis a vis organizational issues, data extraction, integration and analysis and finally taking mutually cohesive decision across the supply chain are the three primary challenges or expectations while applying SCA (Sahay & Ranjan, 2008). But it is also worthy to note that the value addition using IT cannot be realized unless it is aligned with corporate strategy and supported by adequate managerial skill (Dong et al., 2009). An analytic supply chain links data management, process management and performance management to seamlessly integrate people, process and organization in an informative congenial environment for fostering real-time collaborative forecasting and planning for meeting various demands in an agile way (Chae & Olson, 2013; Raghu & Vinze, 2007; Lockamy & McCormack, 2004). Waller & Fawcett (2013) highlighted the importance of predictive analytics in designing competitive supply chain structure. According to them, predictive analytics enables the organizations to estimate the level of business process integration, cost and required service level using both quantitative and qualitative analysis.

Trkman et al. (2010) mentioned various potential applications of SCA in the context of SCOR framework which is summarized in Table 2.

**Table2.** Potential applications of SCA

| Stage | Significant application of SCA |
|---|---|
| Planning | Prediction of market requirement of products and services, forecasting, dynamic pricing |
| Sourcing | Supplier selection and evaluation, price negotiation |
| Making | Production planning and control, inventory planning |
| Delivery | Logistics management, location planning, warehouse planning, network planning |

For example, SCA enables the organizations to forecast firm's sales performance and market requirement on a real-time basis for a region or for a specific location. SCA estimates the margins vis a vis optimal supplier mix, uses transportation intelligence system for notifying movement of goods and possible delay, informs arrival of new products. Further applying association rule mining, dynamic market basket analysis is carried out to offer best possible combination of products to the customers. Using predictive maintenance system which operates on vehicle sensor information, the organizations can improve *overall equipment effectiveness* (OEE) and availability of the machines and can maximize shelf life of the business equipment and machines.

Research findings of Gartner highlighted the significance of both predictive and prescriptive analytics in ensuring real-time visibility across the entire supply chain for improving forecasting, planning, collaboration, networking, sourcing, making and delivery (Gartner Research, 2013). Davenport & Harris (2007) put forth three essential attributes of analytics as quantitative techniques, predictive modeling and optimization. According to them, with the help of these attributes organizations can not only increase the profit margin but also grow the ability of resilience while working in an uncertain business environment. Based on their conviction, the authors propounded an elaborated definition of analytics as: "The extensive use of data, statistical and quantitative analysis, explanatory and predictive models, and fact-based management to drive decisions and actions" or in other words "a subset of business intelligence" (Davenport & Harris, 2007).Over a passage of time many researchers and practitioners put across some key considerations for using analytics. According to them, analytics intends to capture information on what has happened, how it has happened and why it has happened from historical data, evidences and reports (what is technically known as *Descriptive analytics*) in order to understand what is happening now, how it is related to past, what may happen (also known as *Predictive analytics*) for finding what is the best or worst possible outcome, what is the best possible way to operate with undermining the constraints (also called the *Prescriptive analytics*) (Davenport et al., 2010; Gorman, 2012). In this context, the study of IBM reported the importance of analyzing unstructured data like social media data besides the aforesaid structured enterprise and transactional data.

However, Waller & Fawcett (2013) mentioned that appropriate data analysis combined with resource deployment and use of tools not only brings about opportunity for the organizations but also imposes challenges. In their study on data science, predictive analytics and Big Data, they felt the importance of skill development in order to reap potential benefits of using Big Data analytics (Waller & Fawcett, 2013). Cooke (2013) advocated that Big Data analytics is one of the top three influential tools which organizations can leverage for improving supply chain performance. For achieving *big impact*, Ittmann (2015) envisaged that success of SCM lies in its evolving nature and imbibing the concepts and practices of Big Data analytics as basic premise of SCM. Thus, adapting Big Data analytics and putting into practice is one of the critical success factors of next generation SCM (Ittmann, 2015). Specially, advent of e-commerce has added more importance on analytics. In this context, the book retail shop example originally illustrated by McAfee & Brynjolfsson (2012) is worth mentioning. In this example, the authors explained the importance of analytics in understanding consumer behavior, their buying pattern and nature of association of purchased item with other items in the store unlike traditional retailer of books. Retailers can also track the navigation of the buyers or potential customers, capture and analyze their opinions or inquiries, impact of promotional activities and reviews, predict the future requirements and also can understand competitive positions of the products and brands on real time basis using advanced algorithms and data mining tools and techniques. Success of world retail giant like Amazon in this regard clearly supports the growing importance of analytics (McAfee & Brynjolfsson, 2012). Thus it is evident from the above example that Big Data analytics has a strong footfall on designing of competitive supply chain structure for coming decades.

3. **Big Data analytics in the context of SCM**

*Big Data* is basically a voluminous amount of data (both structured and unstructured) of the order of Zettabyte and more which can't be managed by a conventional database management system. Sen (2015) included the improved Gartner definition which expressed that, "Big Data (Data Intensive) Technologies are targeting to process high-volume, high-velocity, high-variety data (sets/assets) to extract intended data value and ensure high-veracity of original data and obtained information that demand cost-effective, innovative forms of data and information processing (analytics) for enhanced insight, decision making, and processes control; all of those demand (should be supported by) new data models (supporting all data states and stages during the whole data lifecycle) and new infrastructure services and tools that allows also obtaining (and processing data) from a variety of sources (including sensor networks) and delivering data in a variety of forms to different data and information consumers and devices".

According to Demchenko et al. (2013b), Big Data is explained on five 'V' dimensions such as: (i) *volume* (amount of data being generated), (ii) *velocity* (The speed at which data being generated, accumulated, retrieved and processed), (iii) *variety* (depicts diversified nature of data being generated), (iv) *value* (intrinsic usefulness of data being generated) and (v) *veracity* (inherent correctness subject to variation). In addition to these, two more dimensions such as *data dynamicity* and *linkage* are included to take different changes of data over the life cycle and their interrelation into account. Sathi (2012) interpreted velocity in terms of throughput and latency. For example, social media data streams produce large volume of data at a speed of the order of TB per day. Velocity is of special interest for the marketers now a day. In order to formulate effective promotional strategy and understand the nature of demand, capturing and analysis of real-time social media data is of great importance (Sathi, 2012).

Big Data is originated mainly from organizational systems and transactions (e.g., SCM, ERP, CRM, e-business), machines (e.g., sensors, smart meters, smart cards, scanners, RFID) and media (e.g., Twitter, Facebook, social blogs, web). The nature of Big Data can be structured (data generated from formal structure like records, files, docs (e.g., XLS, PDF, CSV), tables etc. and captured through traditional systems like *online transaction processing* or OLTP), semi-structured (data not formal in nature but contains distinct semantic elements) and unstructured (data with no identifiable formal structure; e.g., text documents, e-mails, blogs, clickstream data, audio, video, image, public web contents etc.) Experts believe that, in the coming decade, major proportion of the total data being generated will be contributed by unstructured data. Business insights from structured data can be obtained through conventional analytical tools like *online analytical processing* (OLAP), data mining and query processing applications. However, for semi-structured or unstructured data, business intelligence is derived using NoSQL, distributed systems like Hadoop MapReduce Analytics, Massive Parallel Processing (MPP), in-memory analytics and other real time analytics. In our previous work we summarized various characteristics of Big Data in the context of SCM which is given in Table 3 (Biswas & Sen, 2016).

Chen et al. (2012) classified the evolution of *business intelligence and analytics* (BI & A) into three levels. Level 1 (BI & A 1.0) largely operates on structured data using traditional DBMS system. Insight is derived using analytical tools like data warehousing, OLAP, data mining etc. Level 2 (BI & A 2.0) is featured by web and unstructured data. Advanced analytics like web analytics, text mining, sentiment analytics etc. are used to draw intelligence. Level 3 (BI & A 3.0) architecture is characterized by the prevalence of mobile and sensor-based contents and it utilizes tools for location-awareness analysis, person-centered analysis, context-relevant analysis, and mobile visualization and *human computer interaction* (HCI). In true sense, Big Data analytics plays important role for level 2 and 3. Unlike traditional analytics which mainly emphasizes on descriptive and diagnosis analytics, Big Data analytics focuses on predictive analytics using data science. Big Data analytics architecture is primarily influenced by the factors like volume, data sources, latency, throughput, data quality, security and privacy, cost etc. (Chen et al., 2012). Leveling et al. (2014) expressed their concern that, the challenge lays in analyzing irregular arriving data sources spread over large, complex and distributed supply chains having multi-organizational and multi-tier linkages. This necessitates the design of an integrated cloud based framework to monitor and improve business performance while operating in a highly complex and distributed environment (Leveling et al., 2014).

**Table 3.** Various Characteristics of Big Data in context of supply chain

| Type of data | Supplier | Manufacturing | Delivery | Sales and Customer |
|---|---|---|---|---|
| Volume | More detail around design data for products, type of products, process, order, inventory, lot size, delivery, lead time, shipment and routing, pricing, tax, payment, return/dispose. | Product design, customer requirement (e.g., specification, choice, demand, order, time of delivery, feedback), process metrics (e.g., throughput time, cycle time, % rejection, capability, reliability, maintenance), production planning and scheduling, inventory (e.g., lot size, order, WIP, scrap/disposal, finished goods, raw material), material storage, shipment and routing, vendor data (e.g., vendor list, purchase data, vendor evaluation, lead time), people data (e.g., skill inventory, training data, deployment details), finance data (e.g., wage, conversion cost) | Demand data (e.g., order, variety, forecasting), lead time, delivery schedule, location data, inventory (e.g., stock level, aging data), shipping and routing (e.g., mode of transport, load, network and path), finance data (e.g., pricing, exchange rate, tax, payment), miscellaneous (e.g., weather, social, economic, regional data), customer data (e.g., choice, feedback), manufacturing data (e.g., inventory status, production plan and schedule, product details), sales data (e.g., promotion, POS data), return/dispose | Point of Sales (POS) data, order status, demand data, customer data (e.g., product, quantity, delivery, lead time, sentiments, feedback, new product, profile, choice, purchase pattern), promotion, finance data (e.g., payment, pricing, discount, exchange), shipment and routing, return/dispose |
| Velocity | Hourly, daily, weekly, monthly, yearly | Hourly, daily, weekly, monthly, yearly | Real time, hourly, daily, weekly, monthly, yearly | Real time, hourly, daily, weekly, monthly, yearly |
| Variety | Various database, web, audio (verbal/telephonic), E-mail, physical document, sensor data, RFID data | Physical document, sensor data, RFID data, camera, various chips, web data, E-mail | Physical document, sensor data, RFID data, E-mail, various database, web, audio (verbal/telephonic) | Physical document, sensor data, RFID data, E-mail, various database, web, audio (verbal/telephonic) |
| Value | New product development, production planning and scheduling, Optimal lot size and inventory planning, shipping and routing, disposal/ recycle | Optimal lot size and inventory planning, product decision, process selection, execution and control, production planning and scheduling, supplier selection, optimizing delivery lead time, routing decision, remanufacture/recycle/disposal | Transportation and network planning, store planning, inventory planning, customer analysis | Predictive demand modelling, customer analysis, network planning, market basket planning, assortment planning, branding and promotion |
| Veracity | Multiple data sources, different formats, lack of reliability in some data sources, presence of noise in the network communication. | | | |
| Analytics | Association rule mining, optimization, network planning, logistics and supply chain planning, Sentiment analytics, stock planning | Optimization, operations research, assignment and schedule planning, new product development, inventory planning, distribution and warehouse planning, Sentiment analytics, forecasting, predictive demand modelling | Logistics and distribution planning, network planning, retailer selection, Sentiment analytics, forecasting | Sentiment analytics, market basket analysis, forecasting, product shelf layout planning |

Note: As far as suppliers are concerned, 5 V's might not change to a great extent but, in case of products supplied by the suppliers, significant variation may occur.

Over last few years, Cloud Computing and Internet of Things (IoT) have drawn attention of researchers and practitioners in the domain of SCM. In the quest to interlace virtual and real world for ensuring real time supply chain operations, Yan (2015) proposed a cloud based integrated framework for sharing information on real time basis to infuse agility in the system and fostering flexible collaboration and integration. In this context, the author pointed out the inefficiencies of Electronic Data Interchange (EDI) and Enterprise Resource Planning (ERP) systems in terms of higher cost, complexity and less flexibility. In their work, the authors emphasized on 'system architecture, virtualization, and servicization technologies' for setting the foundation for information sharing, flexible collaboration and agility (Yan, 2015).We propose a cloud based architecture of a supply chain in the parlance of Big Data which is shown in Figure 2 (Biswas & Sen, 2016).

In our earlier work, we proposed a cloud-based architecture for a supply chain system in the context of Big Data ecosystem (Biswas & Sen, 2016). Figure 2 depicts the architecture.

The architecture presented in Figure 2 comprises six components: (i) sensor and other data acquisition devices, (ii) cloud infrastructure, (iii) data bus, (iv) data storage and management system, (v) data analytics engine, (vi) data visualization and rendering system.

The *sensor and other data acquisition system* includes all objects and devices that receives raw input data at various points of a supply chain. Sensors, actuators, RFID tag enabled objects, camera connected objects, and the objects which are addressable in the paradigm of Internet of Things (i.e. which have IPv6 addresses) serve as the data source (Bandyopadhyay & Sen, 2011). The data generated from the individual objects are passed over the cloud via several gateway devices placed at suitable locations, in the manufacturing and logistic facilities of the suppliers, manufacturers, distributors, retailers and the end customers.

The input data generated in various different formats by different source objects are passed over the *cloud infrastructure*. The cloud is an infrastructure of computing and storage facilities over the Internet that provides a low cost solution to dynamic storage and processing requirements. The raw data generated by the sources are stored in various servers and may also be possibly replicated for ensuring high robustness and increased availability of the system.

The system *data bus* receives data from the cloud. Data that needs real time processing and analysis are routed through the cloud to the data bus in such a manner that the delay encountered in communication is as low as possible. Other data may be transferred only when they need to processed and analyzed. This is taken care of by the routing protocol in-built in the data communication system. The data bus design takes into consideration the storage capacity requirement of the bus so that a large volume of data might be stored for real-time data processing and analytics.

The *data storage and management system* receives that data from the data bus for efficient storage and retrieval of data. The database management system has all capabilities including indexing, buffering and real-time query optimization. In summary, the main functionality of the data storage and management system is to convert the raw input data into a form that can be very efficiently processed by the analytics engine.

The *data analytics engine* is the core module of the entire architecture. The data analytics engine includes smart processing algorithms for efficient extraction of meaningful and valuable information from the vast volume of the raw streaming or static data. This knowledge base system in the analytics engine enables it to learn from the existing rules in the current systems and build new rules. These new rules enable the analytics engine to take intelligent decision under uncertain situations thereby achieving the goal of a smart analytics under uncertainty or in absence of crisp raw input data.

The *data visualization and rendering system* makes a visual depiction of the analytics results so that correct decision can be taken promptly and effectively. Most of the time the decisions may be taken by the system based on its current understanding and knowledge of rules. However, the user can intervene, if necessary, and overrule the decision taken by the system.

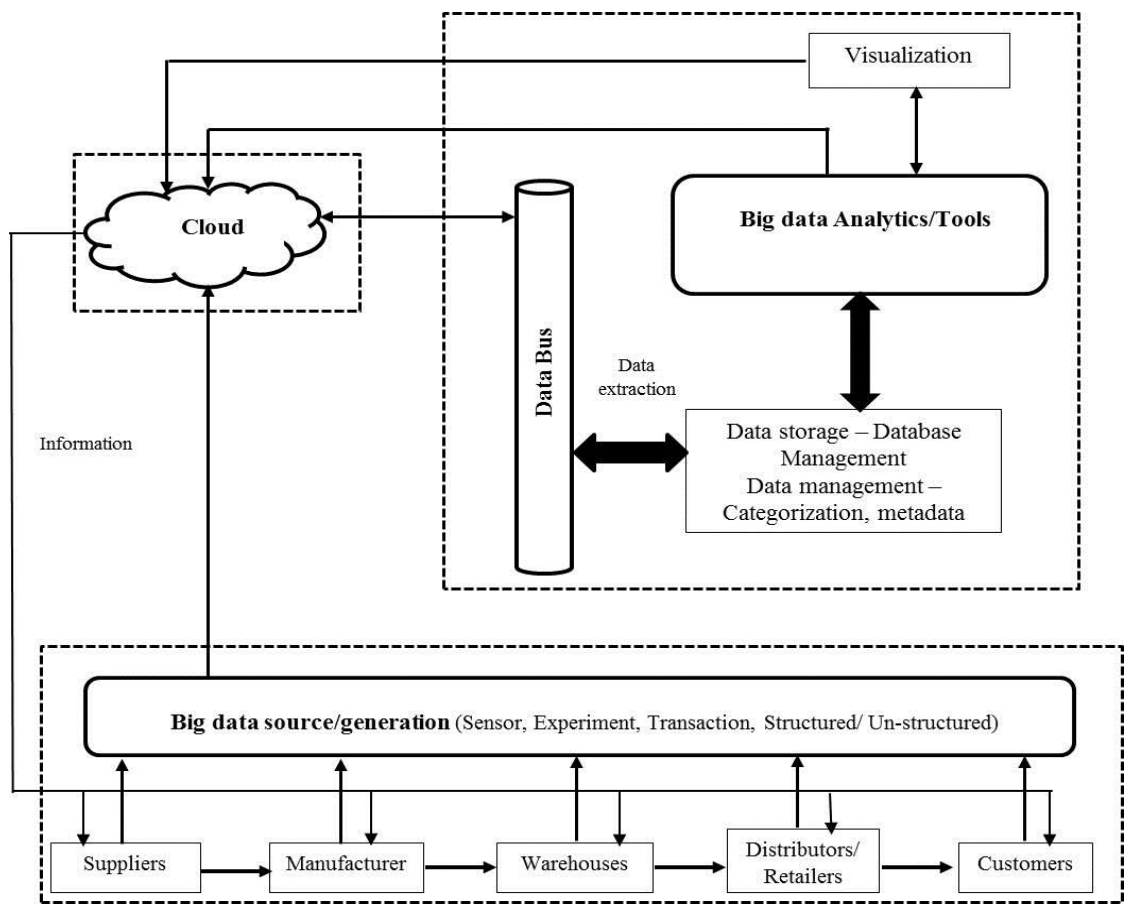

**Figure 2.** Big Data driven next generation supply chain structure

Synchronization of all the components in the architecture is of critical importance so that high availability, robustness, and interoperability of the sub-systems can be achieved. High availability of the system may be achieved by using replications of the components in the system. However, replications involve high cost. System robustness and interoperability among the components can be increased by using state of the art hardware, which too involves higher cost. Hence, a trade-off between operational effectiveness and cost of operations is an important design consideration of the system design.

## 4. Big Data architecture (BDA) of an integrated supply chain

According to IEEE Standard 1471-2000, "Architecture is the fundamental organization of a system embodied in its components, their relationships to each other and to the environment and the principles guiding its design and evolution" (Hillard, 2000).

In this section, we briefly describe a generic Big Data architecture which has been proposed by Chan (2013). This architecture provides a lucid yet a comprehensive description of all the components and their interactions in a typical Big Data system. We then propose a Big Data system for SCM data analytics and management that is capable of handling the volume, complexity and real-time requirement of a typical real-world SCM system.

### 4.1 A Typical Big Data Architecture (Chan, 2013)

Chan (2013) identifies the nature, characteristics and potential applications of Big Data and proposed architecture for Big Data analytics. In essence, the architecture is based on client-server protocol.

In the client side, the author proposes an architecture that consists of NoSQL databases, distributed file systems and a distributed processing framework. A NoSQL database is essentially a non-relational, non-SQL-based data base. However, like a traditional relational database, it stores records in key-value pairs and work very efficiently with unrelated data. NoSQL databases are adaptable to distributed systems and they are highly scalable that makes these databases ideal for Big Data applications. In the client side, below the NoSQL layer, Chan (2013) proposes a scalable distributed file system that is capable of handling a large volume of data, and a distributed processing framework that is responsible for distribution of computations over large server clusters. Hadoop is a very popular distributed file system which can be ideally deployed at this layer. The two critical components in Hadoop that enables it process large data sets across clusters of computers in distributed fashion are: (i) *Hadoop Distributed File System* (HDFS) and MapReduce. While HDFS serves as a storage system that distributes data files over a large number server clusters, the MapReduce is a distributed processing system that processes file in a parallel processing environment.

In server architecture, as proposed by Chan (2013), consists of parallel computing platforms that can handle large volume of data processing at an extremely fast rate. Hadoop architecture includes client machines and a cluster of loosely coupled servers that serve as HDFS and MapReduce data processing core. The client machines load input data into the cluster, submit MapReduce processing jobs, and retrieve the processed output from the server cluster when the processing is complete. The HDFS nodes contain the Name Nodes which are responsible for maintaining the directory of all files in the HDFS file system. The MapReduce nodes, on the other hand, consist of Job Trackers that assign processing tasks to slave nodes. The Job Tracker module identifies the Name Node so as to determine the location of the Data Node that contains the data, and then assigns the task to Task Tracker module residing in the same node, which finally executes the job. The HBase module, built on top of HDFS enables to carry out a very fast record lookups and updates.

Having discussed the client and server architecture of the Big Data system, Chan (2013) proposes analytics architecture over it. The proposed architecture is depicted in Figure 3. The architecture conforms to the *business intelligence and analytics* 3.0 (BI&A 3.0) standard as classified by Chen et al.(2012).

In Figure 3, the Big Data analytics system has the capability to capture structured data through various data sources including OLTP systems, legacy systems and external systems. The raw input data then passes through the *extract transform and load* (ETL) process from the source systems to the data warehouse. The information residing on the data warehouse is used with *business intelligence* (BI) analytics tools such as OLAP for enhancing business operations and decision processes. Unstructured and semi-structured input data sources in Big Data systems can be of numerous types such as: clickstreams, social media, satellites, web logs, sensors, mobile devices, machine-to-machine, geospatial devices etc.

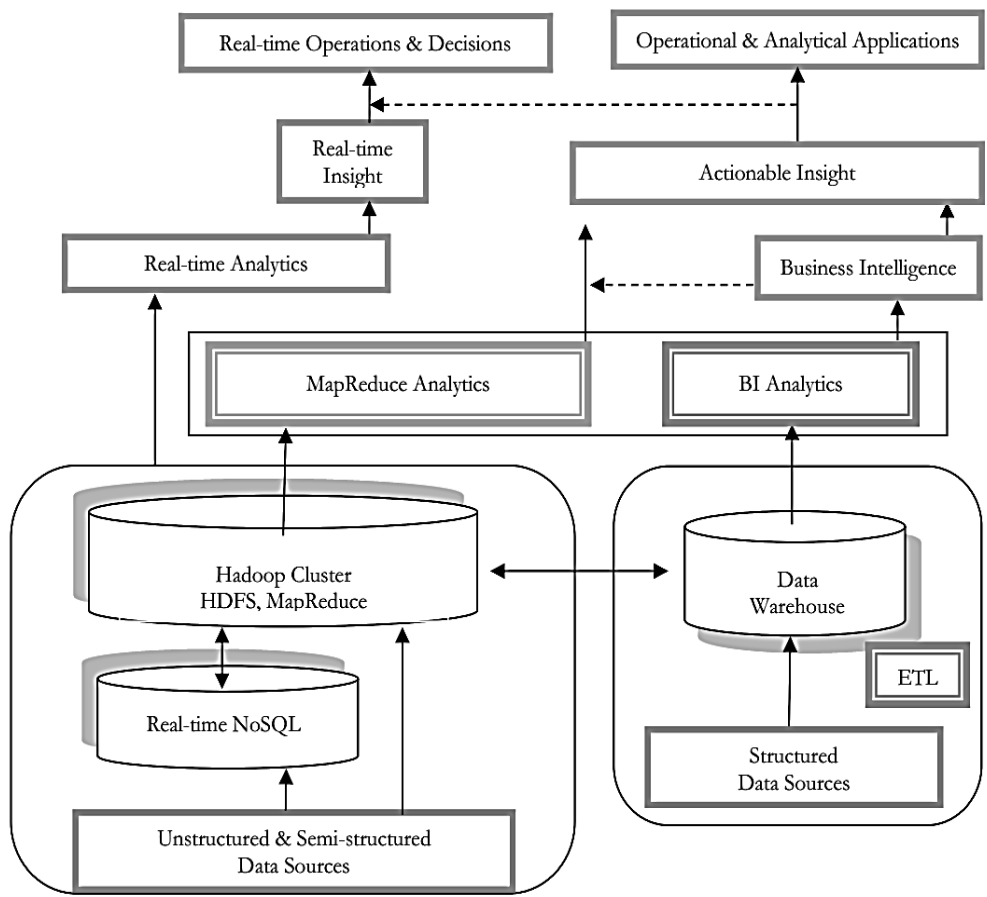

**Figure 3**. Architecture of Big Data analytics (Adapted from Chan, 2013)

Unstructured data are loaded into the HDFS cluster. HDFS cluster provides input to the MapReduce system for distributed processing of the data across Hadoop cluster. MapReduce is also capable of executing batch analytics on humongous volume of unstructured data. The output of analytics results into actionable insights that are consumed by operational and analytics applications. HDFS and MapReduce make the architecture extremely scalable and efficient both from data storage and management and data processing perspectives.

**4.2 A Proposed Big Data Architecture for SCM**

Following the architecture proposed in (Chan, 2013) we propose a Big Data architecture and an analytics framework for SCM applications. The schematic diagram of the architecture is depicted in Figure 4. The various input data sources in the supply chain are represented by the entities at the bottom most layer. These entities are: suppliers, manufacturers, warehouses, distributors/retailers and the customers. These entities provide the input data in both structured and unstructured formats. While data generated and retrieved from traditional databases such as relational databases are structured data, the input data received from various sensors, RFID tags etc. are all unstructured in nature. The agglomeration of these humongous volumes of data results into the generation of Big Data in the system. The Big Data is then supplied as input to the Big Data architecture. While the structured data is extracted by ETL mechanisms and is populated into a data warehouse, the unstructured counterpart is taken care of by the HDFS and MapReduce systems of the Hadoop cluster and is also stored in the NoSQL database. An *operational data store* (ODS) is also deployed after the ETL operation for the structured data inputs before they are populated into the data warehouse. The ODS is a database that is capable of integrating data from multiple sources so that various additional operations can be carried out on the data. While in the ODS, data can be pre-processed, scrubbed, resolved for redundancy and checked for integrity and compliance with the business rules. Structured input data used in the current operation can be housed in the ODS before it is transferred to the data warehouse for persistent storage and archiving. The data in the data warehouse is then accessed by a *real time intelligence* (RTI) system. RTI is an approach to data analytics that users to get real-time data by directly accessing operational systems or feeding business transactions into a real-time data warehouse and *business intelligence* (BI) system. The technologies that enable real-time RTI include data virtualization, data federation, *enterprise information integration* (EII), *enterprise application integration* (EAI) and *service oriented architecture* (SOA). RTI supports instant decision-making since it uses complex event processing tools to analyze data streams in real-time and either triggers automated actions or alerts the users to patterns and trends. The output of the RTI may be directly fed into the analytics applications for the user to visualize the results of analytics in real-time. However, for non-real-time analytics the output of RTI may be fed into a *dimensional data store* (DDS). A DDS is a database that stores the data warehouse data or the output of the RTI module in a different form than the format of OLTP in traditional relational database systems. The reasons for getting the data from the source data warehouse into the DDS and then querying the DDS instead of querying the source data warehouse directly is that in a DDS the data is arranged in a dimensional format that is more suitable for analytics engine. Moreover, when the ETL system loads the data into the DDS, the data quality rules carry out various data quality checks. Bad quality data is fed back into the *data quality* (DQ) database for correction by the source data warehouse. The ETL system is thus managed and orchestrated by a control system, based on the sequence, rules, and logic stored in the metadata of the data warehouse. The metadata is a database containing the summary of the data in the data warehouse and includes information such as the data structure, the data meaning, the data usage, the data quality rules, and other information about the data. The outputs of both RTI module and the DDS module are fed into the Data Mining module that is responsible for finding patterns and relationships in the data that is of interest for the analytics engine. The output of the analytics engine is then suitably presented to the user by rich visualization techniques in the form of reports, charts, graphs etc. In some cases, for patterns and alerts that do not require complex data mining algorithms to identify them, the output of the DDS module may be directly fed into the Analytics engine.

## 5. Protocols for Security in Big Data Systems

With the advent of Big Data, a massive amount of information in a system is available to its user. However, some of the data and information may be private and sensitive in nature and it may not be desirable that all users in the system should be able to get access to that information. Establishing security and ensuring privacy of sensitive data is a vital requirement for any Big Data system.

*Federated Access and delivery Infrastructure* (FADI) has been defined as Layer 5 in the generic SDI Architecture model for e-science (e-SDI) (Demchenko et al., 2013a; Makkes et al., 2013). It includes infrastructure components, including policy and collaborative user groups support functionality. The main component of the FADI standard includes infrastructure components to support inter-cloud federations service such as Cloud Service Brokers, Trust Brokers and Federated Identity Provider. For ensuring security in the system and privacy in data communication, each service/cloud domain contains an *identity provider* (IDP), *authentication, authorization and accounting* (AAA) service agent. These entities typically communicate with their peers in other domains via service gateways.

Privacy of sensitive data is ensured by the use of a *data centric access control* mechanism. Based on the data type and format, two basic types of access control mechanisms are deployed: (i) resource-based access control, (ii) document-based access control. *eXtensible Access Control Markup Language* (XACML) policy language is ideally suited for document-based access control. For the purpose of resource-based access control, native access control mechanisms can be used.

XACML language is very efficient in defining policies for fine granular access control (Demchenko, et al., 2013a). Based on the request context attributes such as subject/user, data identifiers, actions, actions or lifetimes and also on the structured data contents, fine granular access controls are defined. However, there is a disadvantage of XACML. For large documents or complex data structures, XACML policies may lead to a significant computational overhead.

Most of the commercial NoSQL databases for structured data storage have inbuilt security and access control mechanisms. Most of them provide coarse-grain authorization features, both on user management and on protected data granularity like table-level or row-level security.

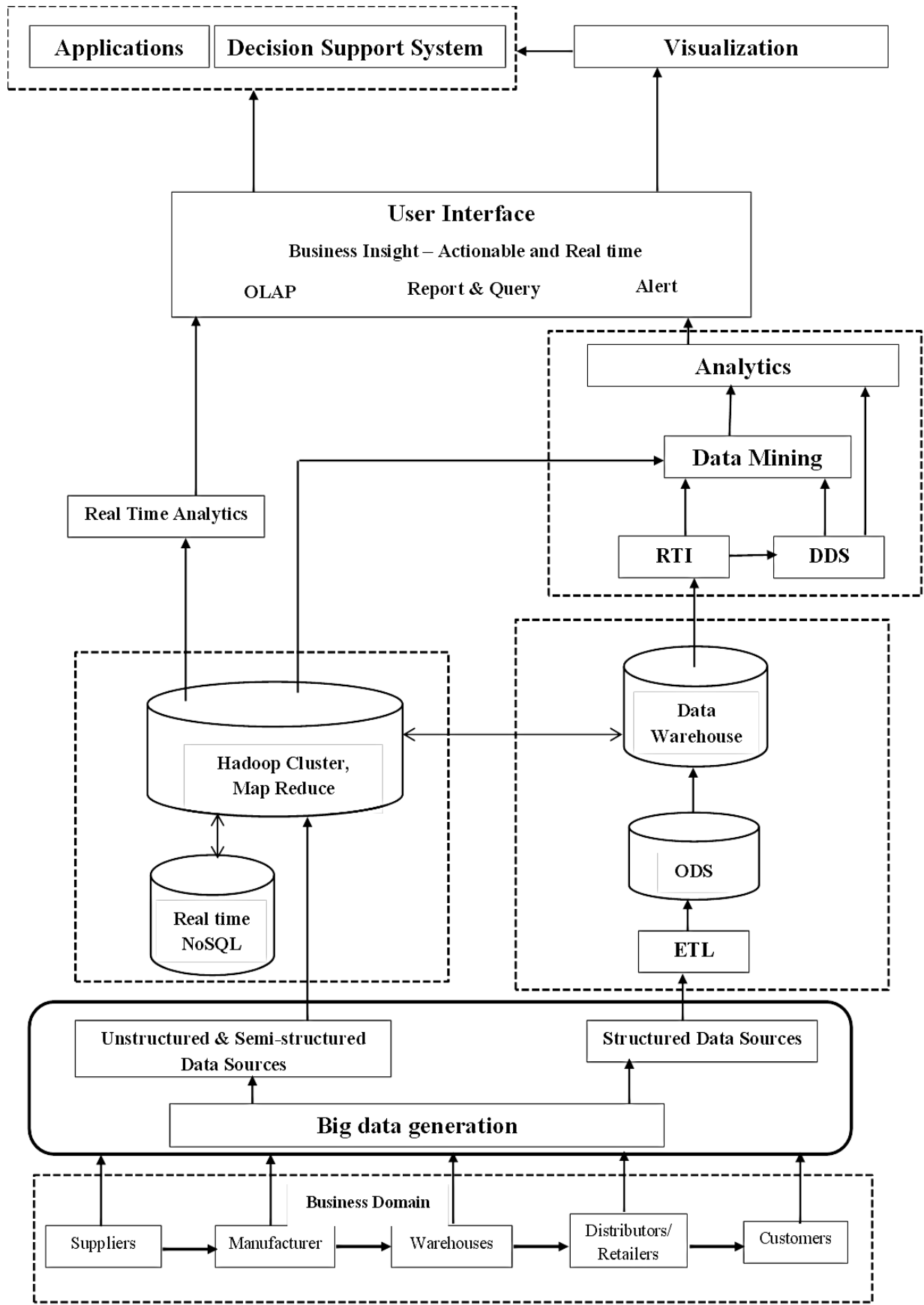

**Figure4.** A proposed architecture of Big Data analytics for SCM

Data-at-rest in remote machines may remain still unprotected even when there are access controls in the databases. Data in remote machines may be protected by using encryption. enhanced access control policies. Encryption enhanced access control mechanisms use attribute-based encryption and allows data decryption only for the targeted subject or attribute owner (Goyal et al., 2006; Chase et al, 2006). Such mechanisms are effective in Big Data use cases for healthcare or targeted broadcast of streaming data in a distributed wireless sensor networks.

Membrey et al. (2012) and Yahalom et al. (1993) have addressed the challenge of creating a trusted distributed computing environment for processing sensitive data in a Big Data ecosystem. The authors have proposed a *dynamic infrastructure trust bootstrapping protocol* (DITBP) that deploys *trusted computing group reference architecture* (TCGRA) and *trusted platform module* (TPM) for establishing trust among the computing entities (Brickell et al., 2004). TPM infrastructure utilizes security at the hardware level by generating a key pair level in the machine hardware. The private key is never revealed and it is used for decryption only when the machine a known and is in a trusted state. The (public, private) key pair is then used for authenticating the machine and then to decrypt the data payload.

## 6. Conclusion

This paper has highlighted the significance of Big Data analytics in the context of SCM. First, the critical role of information in formulating competitive supply chain strategy has been elaborated. Then, with reference to the *supply chain operations reference* (SCOR) framework, the types of data generated across a typical data driven supply chain have been mentioned. A panoramic view of BI and SCA and their importance in the context of supply chain performance have been put forth. Then, the relevance of Big Data in SCA has been discussed. We have also highlighted the growing importance of cloud-based SCM in the current business world and have presented a generic-cloud based architecture for the Big Data centric supply chain operations. Then, an architecture for Big Data analytics in SCM has been proposed building from a generic Big Data architecture existing in the current literature. We have also elaborated the security and privacy requirements in a Big Data system and included a brief discussion on various protocols and mechanisms to enforce these requirements in real-world Big Data systems deployments. The architecture proposed by us is generic nature. The actual design of the protocols for message passing and communication systems development between the modules and the subsystems constitutes our future scope of work. However, it may also be noted that message protocols and communication systems in the overall architecture will vary based on the analytics applications. The real-time systems will impose stringent requirements on delays and reliability while the general applications may specify certain level of tolerance in these parameters. Our systems design will also take into account these characteristics for achieving overall efficiency, reliability and resource optimization.